\documentclass{article}

\usepackage{cite}
\usepackage{amsmath, amssymb, graphics}
\usepackage{graphicx}

\usepackage{textcomp}

\begin{document}


\title{Stimulated optomechanical excitation of surface acoustic waves in a microdevice}
\author{Gaurav Bahl$^{1\ast}$, John Zehnpfennig$^{1,2}$, Matthew Tomes$^1$, Tal Carmon$^1$\\
\\
\footnotesize{$^1$Electrical Engineering and Computer Science, University of Michigan,}\\
\footnotesize{Ann Arbor, Michigan, USA}\\
\footnotesize{$^2$Electrical Engineering and Computer Science, United States Military Academy}\\
\footnotesize{646 Swift Road West Point, New York, USA}\\
\footnotesize{$^\ast$To whom correspondence should be addressed; E-mail: bahlg@umich.edu.}}

\date{}
\maketitle

\begin{abstract}
Stimulated Brillouin interaction between sound and light, known to be the strongest optical nonlinearity common to all amorphous and crystalline dielectrics, has been widely studied in fibers and bulk materials but rarely in optical microresonators. The possibility of experimentally extending this principle to excite mechanical resonances in photonic microsystems, for sensing and frequency reference applications, has remained largely unexplored. The challenge lies in the fact that microresonators inherently have large free spectral range, while the phase matching considerations for the Brillouin process require optical modes of nearby frequencies but with different wavevectors. We rely on high-order transverse optical modes to relax this limitation. Here we report on the experimental excitation of mechanical resonances ranging from 49 to 1400 MHz by using forward Brillouin scattering. These natural mechanical resonances are excited in $\sim$100 {\textmu}m silica microspheres, and are of a surface-acoustic whispering-gallery type. 
\end{abstract}

\begin{figure}[p]  
	\centering
			\includegraphics[width=0.6\hsize, clip=true, trim=0in 2.1in 0in 0in]{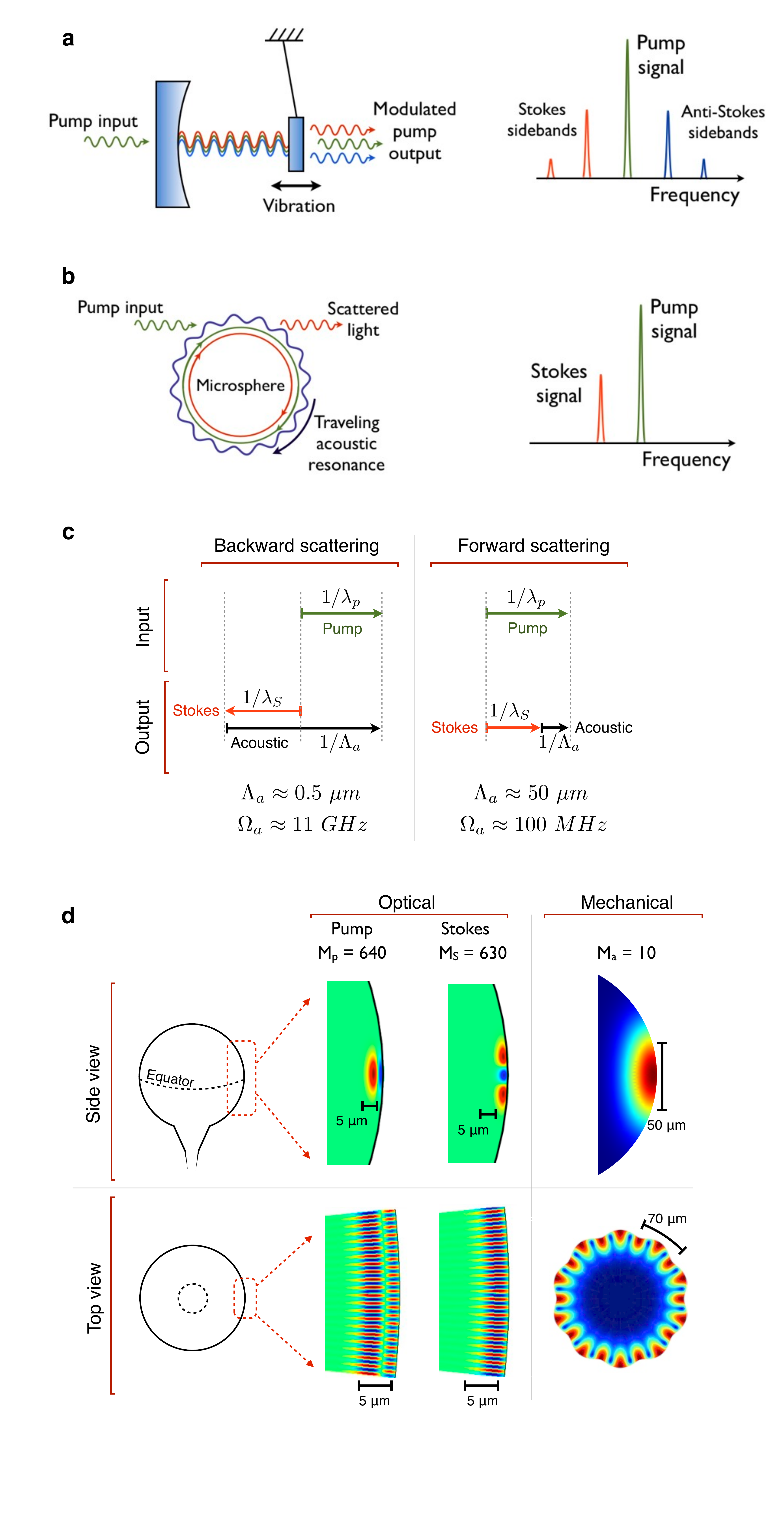}
	\caption{\textbf{Illustration for distinguishing \textit{stimulated} cavity-optomechanics (this work) from \textit{parametric} cavity-optomechanics.} 
	\textbf{(a)} In parametric excitation when one feeds the cavity with the laser right at the optical resonance, the pump is modulated due to a shifting optical resonance, resulting in two optical side bands accompanied by higher harmonics. If the laser is detuned, one can obtain asymmetric spectra to allow net energy transfer from light to the mechanical mode or vice versa.
	\textbf{(b)} In stimulated optomechanics (this work), the pump is scattered to only one Stokes sideband since the acoustic wave acts as a grating continuously moving away from the pump. 
	\textbf{(c)} Wavevector conservation in forward Brillouin scattering leads to a much lower acoustic frequency regime when compared against backward Brillouin scattering.
	\textbf{(d)} Numerically solved optical and mechanical modes for phase-matched stimulated optomechanical interaction on a microsphere resonator, where colors represent electric field of the optical mode and the deformation of the mechanical mode. The mechanical mode (bottom-right) is a Rayleigh-type surface acoustic wave with experimentally observed frequency 57.8 MHz.
	\label{fig:ConventionalVsNew}}
\end{figure}

In the past, interaction of optical and mechanical resonances in cavities arising from modulation of the optical round-trip length has allowed excitation of emission lines 
which were similar to those observed in molecular studies, though only originating from device vibrations \cite{Carmon2005, Rokhsari:05, PhysRevLett.95.033901, CarmonModalSpectroscopy}. Extensive research has since been performed towards extending other atomic-level effects, such as cooling \cite{Arcizet:2006p1092, Gigan:2006p1091, Kleckner:2006p1082} and quantum effects \cite{SchwabQM, PhysRevLett.98.030405, GroundStateCoolingKippen, PhysRevLett.99.093902,Thompson:2008p1083},
to these optomechanical cavities.
The ponderomotive forces used in these experiments \cite{Carmon2005, Rokhsari:05, PhysRevLett.95.033901, Arcizet:2006p1092, Gigan:2006p1091, Kleckner:2006p1082}
 have included centrifugal radiation pressure \cite{Carmon2005}, scattering force \cite{Arcizet:2006p1092, Gigan:2006p1091, Kleckner:2006p1082}, and gradient force \cite{Povinelli:05}. All of these experiments \cite{Carmon2005, Rokhsari:05, PhysRevLett.95.033901, Povinelli:05, Arcizet:2006p1092, Gigan:2006p1091, Kleckner:2006p1082} relied on the \textit{parametric} optomechanical excitation of vibration (Fig.~\ref{fig:ConventionalVsNew}(a)).
However, there is also a different path for cavity optomechanics, of exciting vibration through a \textit{stimulated} process (Fig.~\ref{fig:ConventionalVsNew}(b)), associated with a quantum event involving the emission of a phonon into the mechanical resonance, that we will use here.
Previously, the exploitation of stimulated Brillouin back-scattering \cite{PhysRevLett.12.592, GrudininCaF2lasing} for stimulated photon-phonon emission has been deterred by either zero optical finesse in nanospheres \cite{PhysRevLett.90.255502}, or by negligible mechanical finesse in microspheres \cite{Tomes2009}. 
Fortunately, momentum conservation suggests \cite{MatskoSAWPRL} that reversal of the scattering direction from backward \cite{PhysRevLett.12.592} to forward \cite{Shelby:1985p1169, MatskoSAWPRL}, allows the generation of low-frequency phonons that suffer much less material dissipation \cite{Boyd}, resulting in high finesse acoustic modes. 
In the experiments that we report here, the stimulated process that excites the observed mechanical resonances is based on forward Brillouin scattering \cite{Shelby:1985p1169, MatskoSAWPRL} of photons from acoustic phonons.

Let us consider an optically stimulated traveling acoustic resonance (frequency $\Omega_a$) on a microsphere device with an integer number of acoustical wavelengths ($M_a$) around the circumference (Fig.~\ref{fig:ConventionalVsNew}(d), right). We further consider that this resonance travels uni-directionally along the equator at the speed of sound (Fig.~\ref{fig:ConventionalVsNew}(b), left). 
The process responsible for the generation of this resonance is as follows -- the traveling mechanical mode photoelastically writes an optical grating that is capable of scattering pump light. Since the mechanical mode recedes at the speed of sound, it Doppler red-shifts the scattered light by the correct frequency to enter the Stokes optical mode ($\omega_S = \omega_p - \Omega_a$).
Simultaneously, the remaining pump-mode light along with the Doppler shifted Stokes-mode light can electrostrictively excite vibration at the beat frequency of the two optical modes. If this optically induced stress also travels at the speed of sound, the process described by photoelastic scattering and electrostriction becomes self-sustaining. This intuitive phase-matching condition can be derived analytically by solving the coupled wave equations for mechanical and optical waves to reveal a synchronous solution \cite{Boyd}. 
The solution reveals that the propagation constant of the optical pump $M_p$
(related to azimuthal $e^{i(M_p \cdot \phi -\omega_p \cdot t)}$ propagation)
must equal the sum of the propagation constants of the resulting Stokes and acoustical modes ($M_p = M_S + M_a$). 
This wavevector relation is associated with momentum conservation, as one pump photon is converted into a photon and phonon, whose combined momentum must equal that of the pump photon. This conservation consideration also suggests that low-rate vibrations that are forbidden in backward scattering \cite{GrudininCaF2lasing, Tomes2009} processes (Fig.~\ref{fig:ConventionalVsNew}(c), left) are allowable if forward scattering is utilized instead (Fig.~\ref{fig:ConventionalVsNew}(c), right). Further, it is known that the material dissipation limit for phonons scales inversely with the square of frequency \cite{Boyd}. This suggests that scaling mechanical frequency from 11\,000 MHz (typical of backward scattering in silica) to 50 MHz (this work) can potentially provide a 50\,000-fold reduction in dissipation, transforming the stimulated acoustical wave \cite{PhysRevLett.12.592} into a high finesse mechanical resonance. Phase matching is challenging for forward scattering due to the requirement of having two optical modes with nearby frequencies but with propagation constants differing by $M_a$.  A degree of freedom that can help in achieving close frequencies for the $M_p$, $M_S$ modes is to change the transverse order of the mode till the optical frequencies are properly separated, as described in \cite{Carmon:2007p1167, MatskoSAWPRL}.

Here we experimentally demonstrate the excitation of acoustical modes of a silica microsphere by stimulated forward Brillouin scattering. The actuation mechanism is experimentally verified through observation of the optical signatures of the process. We measure acoustical frequencies ranging from 49 MHz to 1400 MHz. We expect our technique for exciting surface modes to be relevant for applications in surface-wave sensors and in oscillators.

\section*{Results}
\subsection*{Modeling of optical and mechanical modes}
\textit{We calculate} a resonant trio (Fig.~\ref{fig:ConventionalVsNew}(d)) consisting of two optical modes and one mechanical mode, where the optical modes are calculated as shown in \cite{Oxborrow} and the mechanical modes using FEM (See Methods). First, we set momentum to be conserved for the forward scattering process by choosing modes with azimuthal integer number of wavelengths along the circumference which obey $M_p = M_S + M_a$ (e.g.  640=630+10). We then vary the transverse order of the optical modes until $\omega_p = \omega_S + \Omega_a$ is met so that energy is conserved as well \cite{Carmon:2007p1167, MatskoSAWPRL}.
The mechanical mode that we calculate to fit here is a Rayleigh-type 
\cite{Rayleigh_PlaneWaves} 
surface-acoustic-wave (SAW) \cite{SAWDevicesTelecom} as indicated by the fact that particles move elliptically in the equatorial plane and by the fact that upon increasing $M_A$ its speed asymptotically converges to the speed of a Rayleigh SAW on planar substrates as given by the analytical solution \cite{Rayleigh_PlaneWaves}. An acoustic frequency corresponding to the calculation of Fig.~\ref{fig:ConventionalVsNew}(d) will be experimentally measured below in Fig.~\ref{fig:Setup}(b).

\subsection*{Measuring stimulated mechanical vibration}
\begin{figure}[p]  
	\centering
			\includegraphics[width=\hsize]{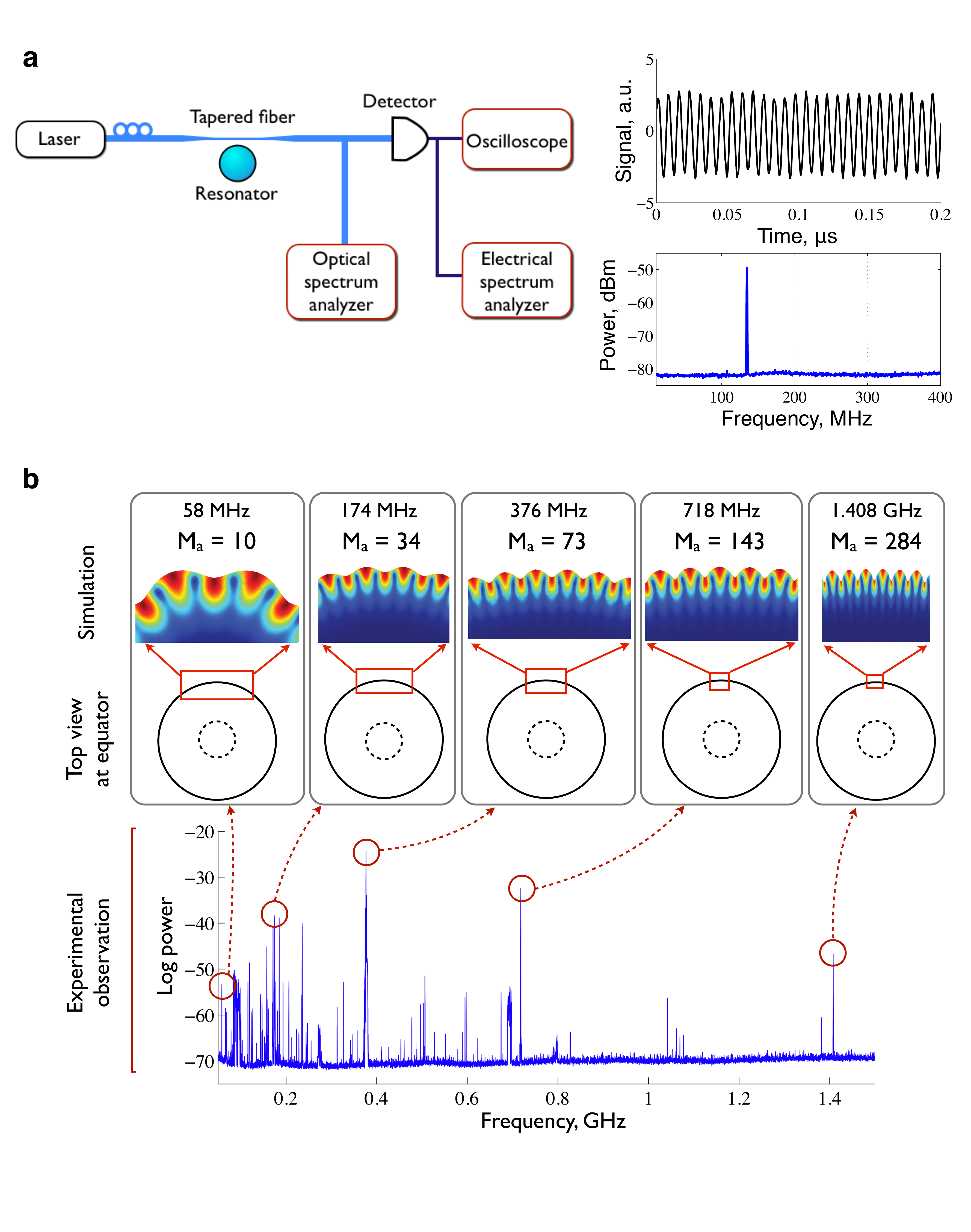}
	\caption{
	\textbf{Experimental setup and experimental modal spectroscopy} -- 
	\textbf{(a)} The pump laser is coupled to the optical microsphere resonator through a tapered fiber. Forward-scattered optical signals and the pump transmission are subject to both optical and electrical spectrum analysis. The experimental data insets show an example 134 MHz mechanical mode observed in electrical domain with an unchanging pump laser.
	\textbf{(b)} Many mechanical modes are observed when the pump laser is scanned slowly from 1520 - 1570 nm. We depict the best-match Rayleigh surface modes for several of the measured frequencies. Colors represent the deformation of the mechanical mode.
	\label{fig:Setup}}
\end{figure}

\textit{Mechanical vibration is stimulated} by pumping the appropriate optical resonance of a microsphere using a tunable pump laser (See Methods, and Fig.~\ref{fig:Setup}(a)). The vibration is measured via the beating of the pump and Stokes modes on a photo detector. Temporal sinusoidal dynamics are observed (Fig.~\ref{fig:Setup}(a), top right) together with the corresponding spectral line (Fig.~\ref{fig:Setup}(a), bottom right) that we measure with an electrical spectrum analyzer. 
We then use an optical spectrum analyzer to verify that this beat indeed originates from a Stokes line that is separated from the pump by the acoustical frequency. Such correlation is shown in Fig.~\ref{fig:TheStory}. Additionally, a broadband spectrum analyzer is used to verify that four-wave mixing \cite{PhysRevLett.93.083904} or Raman lasing \cite{Spillane:2002p1081} do not interfere with our measurements (See Methods).

\textit{Modal spectroscopy measurement} is performed by slowly tuning the laser through optical resonances while recording the vibration frequencies excited through resonant trios similar to what was calculated in Fig.~\ref{fig:ConventionalVsNew}(d), but having different modal orders. In the example shown in Fig.~\ref{fig:Setup}(b), we observed mechanical modes ranging from 58 MHz - 1.4 GHz in a single sphere with the laser being tuned from 1520 - 1570 nm. We depict several of these experimentally observed frequencies and provide their closest mechanical Rayleigh modes.

\begin{figure}[htbp]  
	\centering
			      \includegraphics[width=0.8\hsize]{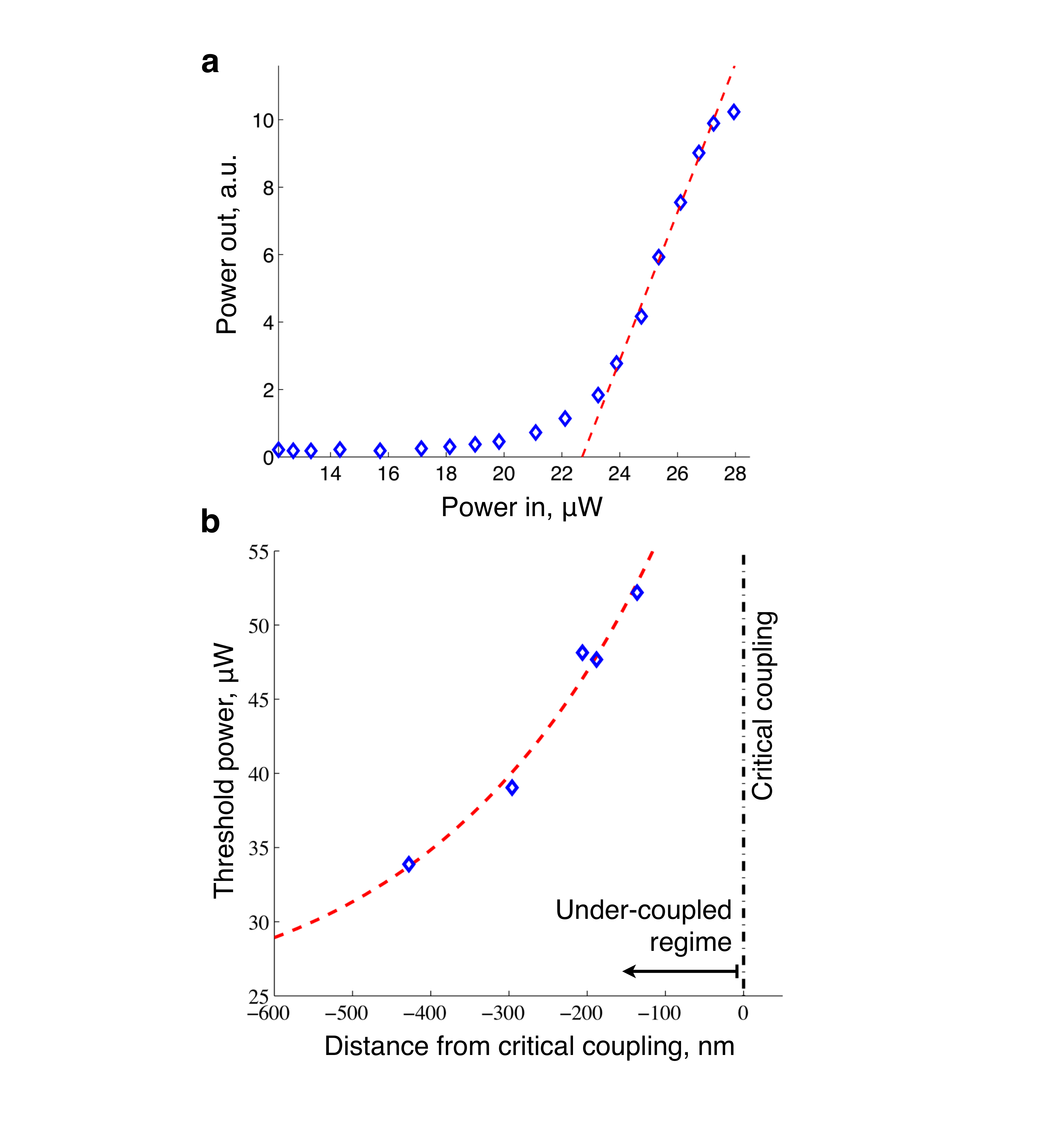}   
	\caption{
	\textbf{Experimental observation of threshold power required for vibration.}
	\textbf{(a)} Power in vs. power out curve obtained for a 175.36 MHz mechanical vibration.
	The linear fit line indicates $P_{\textrm{threshold}} \approx 22.5~{\mu}W$.  
	\textbf{(b)} Under-coupling reduces the threshold power. This data is obtained for a 134 MHz mechanical mode. Dashed line is a fit to the theory presented in Methods.
	\label{fig:Thresholds}}
\end{figure}

\textit{The vibration threshold} is measured to be 22.5 ${\mu}W$ (Fig.~\ref{fig:Thresholds}(a)) for a 320 ${\mu}m$ diameter sphere with quality factor $\sim$400 million, pumped at 1.5 ${\mu}m$ wavelength, mechanically resonating at 175.36 MHz. As with other stimulated processes including Raman \cite{Spillane:2002p1081} and Erbium \cite{min:181109} microlasers, threshold power depends on the coupling distance between taper and resonator as measured in Fig.~\ref{fig:Thresholds}(b) and fitted to what is predicted by coupled-mode-theory analysis \cite{HausCoupledTheory} (see Methods). Qualitatively and to a certain limit, this threshold lowering behavior is due to under-coupling reducing Stokes transmission out of the cavity and leaving more of it inside the resonator to provide electrostrictive forces that excite vibration.

\begin{figure}[p]  
	\centering
			\includegraphics[width=0.8\hsize]{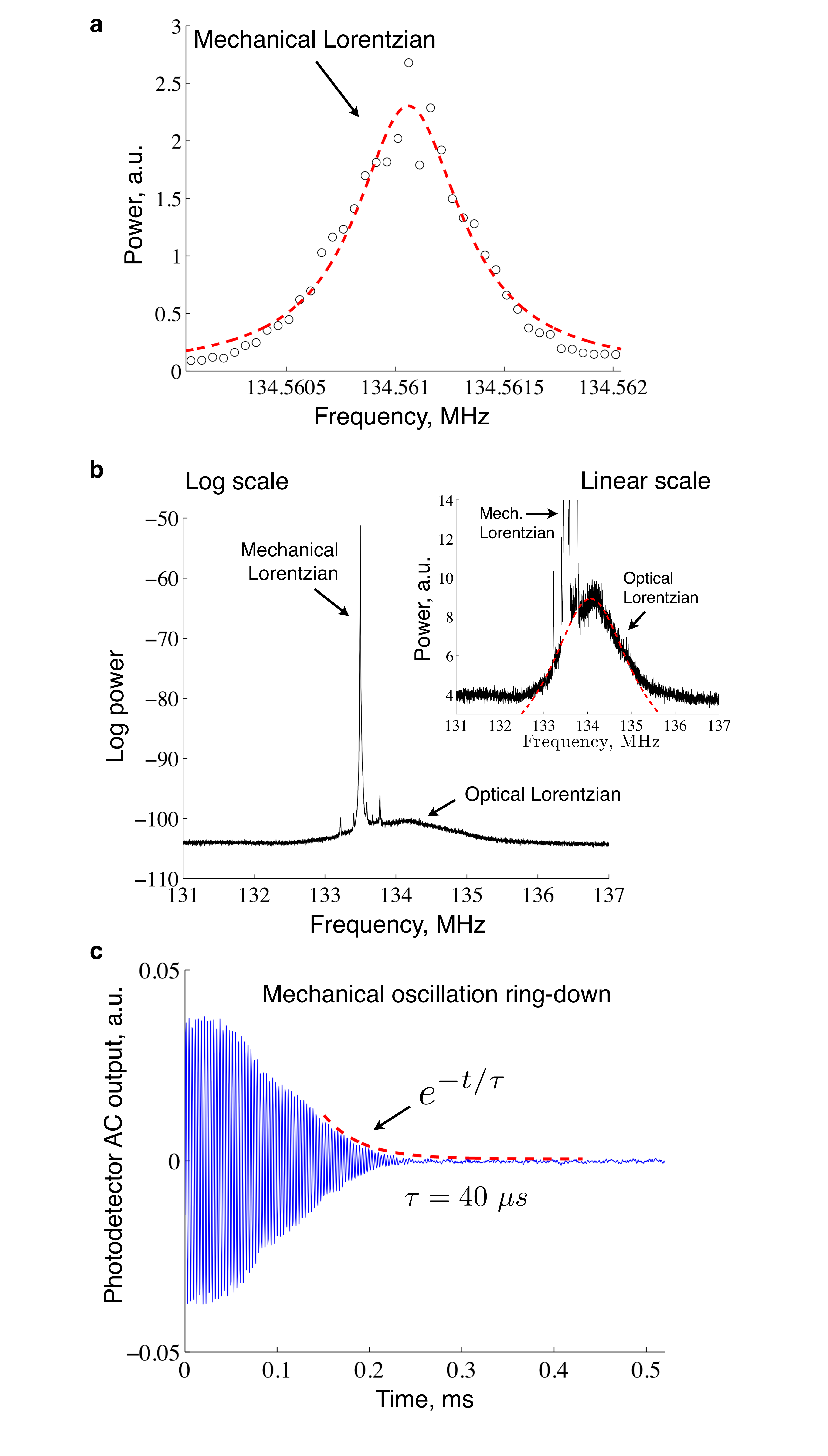}
	\caption{
	\textbf{Linewidth and mechanical mode characterization} --
	\textbf{(a)} Sample beat linewidth captured for a $134$ MHz mode.
	The plotted Lorentzian guide indicates a FWHM linewidth of 580 Hz. 
	\textbf{(b)} The mechanical signal is superimposed on a low power MHz-width Lorentzian that originates from the optical resonance. 
	\textbf{(c)} Vibration ring-down experiment for determining the mechanical quality factor of a 175 MHz mechanical mode on a sphere of radius R = 162 $\mu$m.
	\label{fig:Oscillator}}
\end{figure}

\textit{Measuring the fine structure of the beat-note spectrum} reveals a narrow and strong Lorentzian ($\sim$580 Hz, Fig.~\ref{fig:Oscillator}(a))  superimposed over a weak and wide Lorentzian ($\sim$1 MHz, Fig.~\ref{fig:Oscillator}(b) inset). The taller Lorentzian originates from the width of the vibration gain spectrum that is expected to be $\sim$2 kHz wide \cite{Boyd}. In resonators such 2 kHz width is expected to narrow \cite{MatskoSAWPRL} while being limited by the Schawlow Townes relation \cite{SchawlowTownes}. The broad Lorentzian originates from the optical resonance width. 
It is worth mentioning that the stimulated process gain here is $\sim$1000x narrower than the optical resonance \cite{Boyd}, in contrast with other processes \cite{Spillane:2002p1081, min:181109,Tomes2009,Vahala:2008p1075} where the relation is opposite. 
Interestingly, linewidths like in Fig.~\ref{fig:Oscillator}(a) are related to both the mechanical quality factor and the optical quality factor.

\textit{The mechanical finesse} is estimated by measuring the vibration decay time. Although many additional factors interfere with this measurement, we attempted to perform a ring-down measurement by executing a fast scan of the pump wavelength. Based on several such scans we estimate a phonon lifetime $\tau \approx$ 40 ${\mu}s$, implying a mechanical quality factor of approximately Q~$\approx$~44\,000 (finesse 900) for the 175 MHz mode shown in Fig.~\ref{fig:Oscillator}(c).  This is in comparison with material limited mechanical Q~$\approx$~48\,000 \cite{Boyd}. In our experiments, the Stokes mode is produced at efficiencies of up to 50\%, with input power in the mW range.

\section*{Discussion}

In this work we report for the first time on the experimental stimulated optical excitation of high finesse mechanical resonances in silica microspheres via forward Brillouin scattering, indicating a new mechanism for studies in cavity optomechanics. The stimulated mechanical whispering gallery modes that are generated here are surface acoustic waves of the form typically excited using metal electrodes on piezoelectric substrates \cite{SAWDevicesTelecom}. In our case, however, the electrodes are virtual, and are written by light traveling with the mechanical mode. 

The dissipation suffered by phonons at these mechanical resonances is near the material limit, since they are whispering-gallery modes that propagate azimuthally just like the optical whispering gallery modes that excite them. Such concentration of vibration at the sphere surface removes the necessity for notch supports and mechanical quarter-wavelength leakage isolators for the resonator, simplifying the mechanical design of the system. This is unlike the case of breathing modes \cite{CarmonModalSpectroscopy} that have propagation components in the direction of the mechanical supporting structures.  
Additionally, the mechanism of electrostriction that supports the stimulated Brillouin process, is common to all dielectrics inclusive of every amorphous or crystalline optical material. Such Brillouin amplification benefits from having the highest gain of all the optical nonlinearities \cite{Boyd}. In silica, for instance, Brillouin gain is more than two orders of magnitude higher than Raman gain \cite{Boyd}.

Microresonators that are presently at the core of many experiments in optics for enhancing electromagnetic fields \cite{VahalaMicrocavities, ChangMicrocavities} 
are shown here to allow stimulated optical excitation of high finesse whispering gallery mechanical resonances. 
The narrow oscillation linewidth and the fact that the mechanical mode is concentrated to within one acoustical wavelength from the interface with the environment make this device attractive for sensing changes in the environment \cite{SAWSensors}.
Further, these mechanical resonances see dissipation near the material-property limit, and are consequently very useful in oscillator applications as well.

\section*{Methods}

\subsection*{Numerical modeling of optical and mechanical modes}
The mechanical modes that we experimentally observed were identified using 3D finite element modeling in COMSOL Multiphysics \cite{Comsol}. 
The existence of pairs of optical modes that satisfy phase match to excite these mechanical modes was confirmed using the numerical method described in \cite{Oxborrow}. Gross tuning of phase match was achieved by discretely changing the transverse order of the optical modes. Fine tuning was then performed by modifying the ellipticity of the sphere till exact phase match was reached. Shape and size of the calculated device were kept within the measurement error using our microscope.

\subsection*{Experimental setup}
Our experimental setup (Fig.~\ref{fig:Setup}(a)) consists of a silica microsphere resonator \cite{GorodetskyMicrospheres} (typical optical $Q > 10^8$) evanescently coupled to a telecom-wavelength laser ($\sim$1550 nm) through a tapered optical fiber \cite{Cai:01, Spillane:2003p1318, Knight:1997p1319}. We tune the pump laser to one of the optical resonances, and via the other end of the taper, measure the output of the partially transmitted pump along with the Stokes mode that vibration scatters. Optical spectrum analyzers are used to verify optical signals, while photodetectors provide the electrical beat signal. We emphasize that vibration is stimulated with a continuous-in-time single frequency laser, and that no external oscillator, controller, or feedback is used.

\subsection*{Verification of optical signals} 

\begin{figure}[p]  
	\centering%
			\includegraphics[width=0.8\hsize]{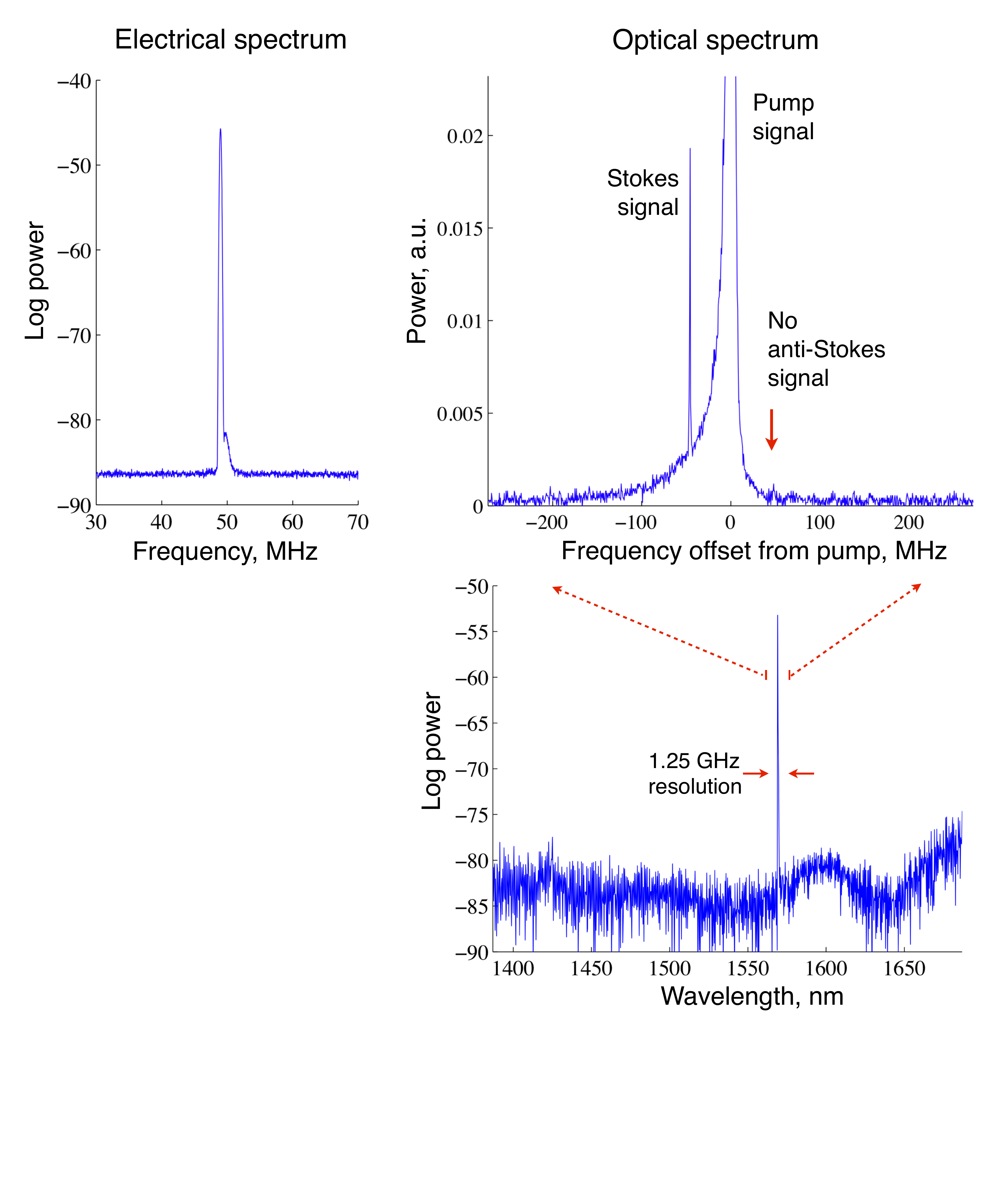}
	\caption{
	\textbf{Electrical and optical spectra for a $\textbf{48.9}$ MHz vibration} in a sphere of radius $R = 162~{\mu}$m with pump wavelength $1567$ nm. The electrical spectrum represents the beat note between the two optical signals. The absence of the anti-Stokes sideband in the optical spectrum as well as the absence of higher harmonics clearly indicates that this is not a mechanical breathe mode. The wide optical spectrum (bottom) confirms that our observation is not disturbed by any other optical lines such as those arising from four wave mixing or Raman scattering.
	\label{fig:TheStory}}
\end{figure}

The presence of a single Stokes side-band to the optical pump is confirmed by optical spectrum analysis through a Thorlabs SA200-14A tunable Fabry-Perot spectrum analyzer (Fig.~\ref{fig:TheStory}, top). 
We confirm that no optical lines other than the stimulated Stokes line exist in our system by using a second OSA (Advantest Q8384 Optical Spectrum analyzer) that overs a broad spectral range (Fig.~\ref{fig:TheStory}, bottom).
Also, as expected, the broad span OSA cannot resolve the two optical signals due to their close spacing.

\subsection*{Derivation for coupling-dependent threshold}

The relationship of coupling threshold to the coupling distance can be derived following the rate equations in \cite{GorodetskyOptimal}.
Starting with the non-depleted pump approximation, a rate equation for the pump mode can be obtained
\begin{align}
	\dot{A_p} = i \cdot C(x) \cdot B_{in} - (\delta_c(x) + \delta_o) \cdot A_p
	\label{eqn:pumprate}
\end{align}
where $A_p$ is the field amplitude of the pump mode, $B_{in}$ is the pump field amplitude in the coupling waveguide, $x$ is taper-sphere distance, $C(x)$ is the taper-sphere distance-dependent coupling coefficient, $\delta_o$ is the intrinsic loss, and $\delta_c(x)$ is the coupling loss. From here, the steady state expression, 
\begin{align}
	A_p = \frac{i \cdot C(x) \cdot B_{in}}{\delta_c(x) + \delta_o}
	\label{eqn:pumpss}
\end{align}
can be obtained.
The Stokes mode rate equation is derived in a manner similar to Eqn.~\ref{eqn:pumprate}.
\begin{align}
	\dot{A_s} = - g \cdot |A_p|^2 \cdot A_s - (\delta_c(x) + \delta_o) \cdot A_s
	\label{eqn:stokesrate}
\end{align}
where $g$ is the process gain and $A_s$ is the Stokes mode field amplitude. By balancing gain and loss terms in Eqn.~\ref{eqn:stokesrate} and incorporating Eqn.~\ref{eqn:pumpss}, the threshold power can be obtained.
\begin{align}
	{B_{in}}^2 = \frac{(\delta_c(x) + \delta_o)^3}{g \cdot C(x)^2}
\end{align}
This is rewritten as 
\begin{align}
	P_{\textrm{threshold}} \propto \frac{ ( T(x)^2 / 2 \tau_0 + \delta_0)^3 }{ g \cdot T(x)/\tau_0 }
\end{align}
where $T(x)$ is the coefficient of transmittance exponentially related to $x$ the taper-sphere distance, and $\tau_0$ is optical round-trip time.

\section*{Acknowledgements}
The authors would like to acknowledge discussions with C.X.~Deng, Y.-S.~Hsiao, and A.B.~Matsko. This work was supported by the DARPA ORCHID program through a grant from AFOSR.

\section*{Author Contributions}
All the authors jointly designed and conceived the experiments. G.B. and J.Z. built the setup, fabricated the devices, and carried out the experiments. M.T., G.B. and J.Z. performed the analytical and numerical calculations. G.B., M.T. and T.C. wrote the paper. T.C. supervised all aspects of this project.


\section*{Competing financial interests}
The authors declare no competing financial interests.


\bibliographystyle{ieeetr}

\end{document}